# BEAM PROFILE MONITORS IN THE NLCTA[*]

C. Nantista, C. Adolphsen, R.L. Brown, R. Fuller and J. Rifkin

*Stanford Linear Accelerator Center, P.O. Box 4349, Stanford, California 94309, USA*

*Abstract*

The transverse current profile in the Next Linear Collider Test Accelerator (NLCTA) electron beam can be monitored at several locations along the beam line by means of profile monitors. These consist of insertable phosphor screens, light collection and transport systems, CID cameras, a frame-grabber, and PC and VAX based image analysis software. In addition to their usefulness in tuning and steering the accelerator, the profile monitors are utilized for emittance measurement. A description of these systems and their performance is presented.

## I. INTRODUCTION

Among the devices which comprise the beam instrumentation system of the Next Linear Collider Test Accelerator [1] at SLAC are eight beam profile monitors: one near the gun, one between the injector sections, one each before, in the middle of, and after the chicane, one after the main accelerator, and one at each of the two dumps. Along with four wire scanners, these profile monitors allow us to view the transverse charge distribution of the electron beam at various positions along the beam line. They are of great utility for monitoring beam shape, verifying transmission, and, in the high dispersion region of the chicane and spectrometer, measuring energy spread in the long bunch train [2],[3]. In addition, they are employed in data acquisition for beam emittance and TWISS parameter measurement.

## II. SCREEN

The screens used in our profile monitors are of the same type used in the SLC [4]. They are fabricated from aluminum, to which is bound a fine-grained surface layer of $Gd_2O_2S:Tb$. This coating phosphoresces with a decay time of about a millisecond where the incident electron beam deposits energy. The light emitted per unit area is proportional to the charge passing through the screen. They are supported on pneumatic vacuum feedthroughs which allow them to be retracted from the beam pipe volume when not in use.

When a screen is inserted, its surface makes an angle of $45°$ with the beam line axis so that it can be viewed through a vertical view port. A pattern of holes drilled in the screen allows for image dimensions to be calibrated. For most screens, this is a rectangular frame with seven holes horizontally and five vertically. The horizontal and projected vertical hole spacing is 1.5 mm. The last three profile monitors have larger screens with a different hole pattern.

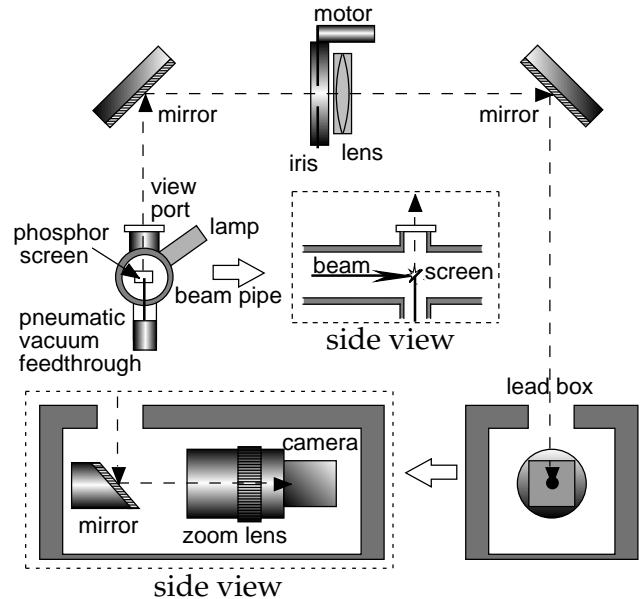

Figure 1. Profile monitor optics diagram.

## III. OPTICS

An optical system is employed to distance the camera from the high radiation levels produced by the beam, particularly when incident on the screen. The layout of our optical elements is illustrated in Figure 1. They include three flat optical mirrors, an adjustable iris, and a three-inch, 350 mm focal length imaging lens. The image magnification is set close to unity to minimize aberration. The camera is placed a couple of feet off to the side of and below the beamline and is shielded by an enclosure of lead bricks supported by a steel frame. The amount of light entering the camera is adjusted by means of a remotely controlled motor connected to the iris. Also remotely controllable is a lamp aimed through a second window in the vacuum chamber. This allows us to illuminate the screen in order to view the hole pattern and calibrate the x and y scales of the received image. Illumination is also needed while manually aligning the optical elements and focusing the image.

## IV. CAMERA

The camera used in our profile monitors is the CID2250D manufactured by CIDTEC, a 30 Hz, solid state, CID (charge injection device), monochrome video camera. The designation "CID" refers to the fact that after each frame is read out from the sensor, the charge on its MOS capacitor array is injected into a substrate. It is rated for fifty times longer life than CCD (charge coupled device) cameras in an ionizing radiation environment and is much more "neutron-hard". The sensor is a 59 square millimeter surface containing $512 \times 512$ pixel capacitors with 0.015 mm spacing. Its sensitivity and range can be adjusted by setting the gain and offset voltage.

---
[*] Work supported by Department of Energy contract DE-AC03-76SF00515.

The camera is fitted with a close-up zoom lens which is focused on the screen image formed several centimeters away. A long, special cable connects the palm-sized camera body to a camera control unit (CCU) outside of the shielded accelerator housing.

## V. FRAME-GRABBER / TIMING SIGNALS

Each of the eight CCU's is connected via two coaxial cables through a multiplexer and some electronics into a frame-grabber in the NLCTA control room. For the latter we employ Data Translation's Fidelity 200 Flexible Frame Processor board installed in an Industrial Computer Source 90 MHz PC. Figure 2 shows a diagram of our profile monitor signal processing system.

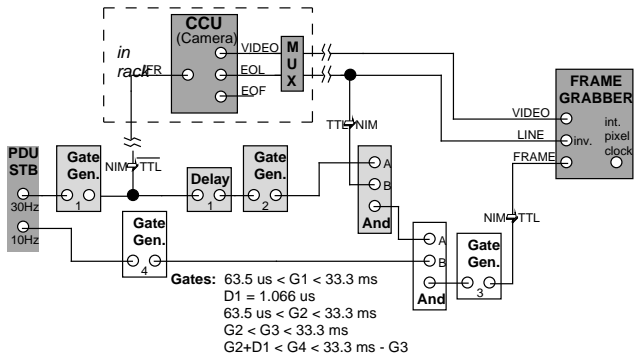

Figure 2. Timing signal electronics.

The frame-grabber operates on three timing signals: pixel clock, end-of-line, and end-of-frame. Using the board's internal oscillator for the pixel clock was found to eliminate a line-to-line image jitter problem. The end-of-line signal, output by the camera control units, is brought in through one of the coax cables. The other cable carries the raw video signal. To avoid stringing more cables than necessary, and because our multiplexer only switches two signals, we worked out a way to run without bringing the CCU's end-of-frame (EOF) signal into the frame-grabber, as described below.

A third coax cable, daisy chained to all of the CCU's from the control room, carries to them a 30 Hz frame-reset signal to synchronize the cameras with the accelerator timing system. The CCU's output end-of-frame signal is locked to its first end-of-line pulse after the leading edge of the frame reset. We can produce the same signal in the control room by combining the frame reset and incoming end-of-line signals in a NIM logic unit. This process is illustrated in Figure 3. A time delay reflectometer was used to determine the round-trip travel time between the control room and the CCU, so that the delayed frame reset pulse would select the proper end-of-line pulse. The necessary components are shown in light gray in the diagram of Figure 2.

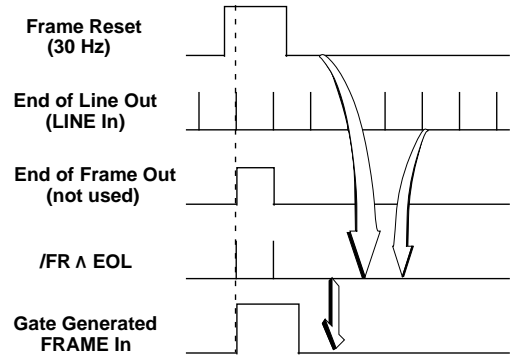

Figure 3. Generation of end-of-frame signal.

The cameras must operate at 30 Hz. However, in NLCTA operation, the beam repetition rate is generally 10 Hz. The resulting strobe effect in the video image can be annoying and gives one only a one third probability of capturing a beam image on each try. The following procedure was therefore incorporated. We take a 10 Hz trigger identical to that driving the electron gun from the same PDU (Programmable Delay Unit) that gives us our 30 Hz trigger. We expand this with a gate generator and combine it in a logic unit with our constructed end-of-frame signal. Since the latter has been slightly delayed, the effect is to pass every third end-of-frame pulse, as illustrated in Figure 4. When the frame-grabber is then given this signal and told to expect 10 Hz video, this is all that it digitizes, despite the fact that the camera is sending thirty frames per second. This is accomplished by the unshaded components in Figure 2.

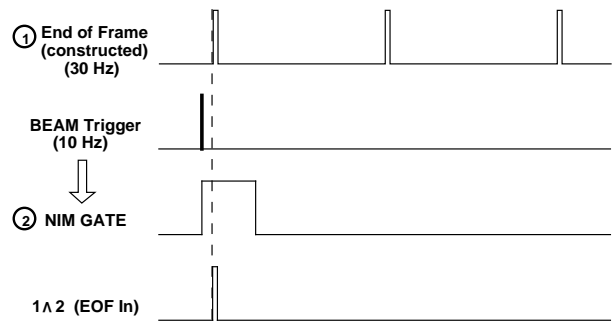

Figure 4. Reduction of end-of-frame frequency to 10 Hz.

## VI. SOFTWARE

In conjunction with the Fidelity 200 frame-grabber card, we use Data Translation's GLOBAL LAB Image

software, running in Microsoft Windows for Workgroups. This image processing and analysis package has many features. It configures the frame-grabber for our camera model. It displays live video or captures single frames, which can be saved as TIFF format image files. It can plot the pixel values along any slice of an image selected by dragging a line. It also allows the user to program scripts to perform desired sequences of operations.

Unfortunately, GLOBAL LAB Image seems not to have been designed with beam physicists in mind. It cannot readily do such things as projections, curve fits, and contour plots. One can expand its capabilities by writing additional C-code. However, for our purposes, it was preferable to FTP beam images to the SLC VAX and analyze them there, when desired, with our own code written in MATLAB. This code subtracts the baseline from the image array, displays 2-D contour plots, projects the pixel values onto the x and/or y axis, and fits gaussian curves to determine the beam sigmas. An example beam spot is shown in Figure 5. Its projections, along with gaussian fits, are shown in Figure 6.

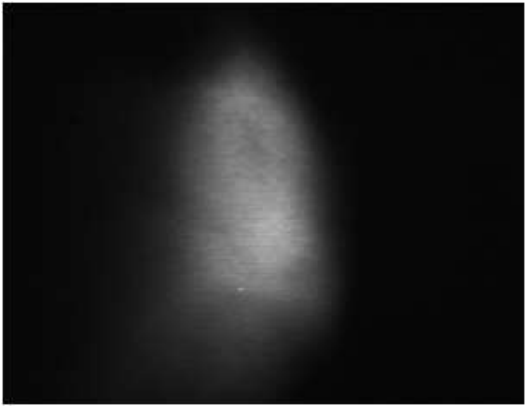

Figure 5. Beamspot image.

## VII. EMITTANCE MEASUREMENT

The beam physics program of the NLCTA makes it desireable to be able to measure the beam emittance, as well as how well matched the beam is to the lattice. This is done with our profile monitors using to the general procedure described by Ross, et al. [5].

To measure horizontal beam phase space, for example, the CORRELATIONS PLOT program in our controls software is used to step a horizontally focusing quadrupole magnet upstream of the chosen profile monitor through seven field strengths about the nominal value. At each step, a trigger is sent to our frame-grabber, and GLOBAL LAB Image stores the beam image to disk.

These images are then transfered from PC to VAX, and their gaussian spacial distribution sigmas in the relevant dimension are determined. These are related to the beam sigma matrix at the screen by $\sigma_{11} = \sigma_x^2$. MATLAB then reads in the quadrupole name and field values from the CORRELATIONS PLOT output and gets the tranfer matrix between the quadrupole and the profile monitor from the online database, as well as the beam energy. A parabola is fit to a plot of $\sigma_x^2$ vs. quadrupole strength. From its three coefficients and the transfer matrix, we can determine the $2 \times 2$ sigma matrix at the quadrupole. The emittance and TWISS parameters then follow from
$\varepsilon = \sqrt{\det \sigma}$, $\beta = \sigma_{11}/\sqrt{\det \sigma}$, and $\alpha = -\sigma_{12}/\sqrt{\det \sigma}$.
Using further transfer matrix information from our model, we can "swim" this beam ellipse upstream and calculate the mismatch at the accelerator input, characterized by

$$B_{\text{mag}} = \frac{1}{2}\left[\frac{\beta^*}{\beta} + \frac{\beta}{\beta^*} + \left(\alpha\sqrt{\frac{\beta^*}{\beta}} - \alpha^*\sqrt{\frac{\beta}{\beta^*}}\right)^2\right],$$

where the asterisked parameters are the measured values and the others the design values.

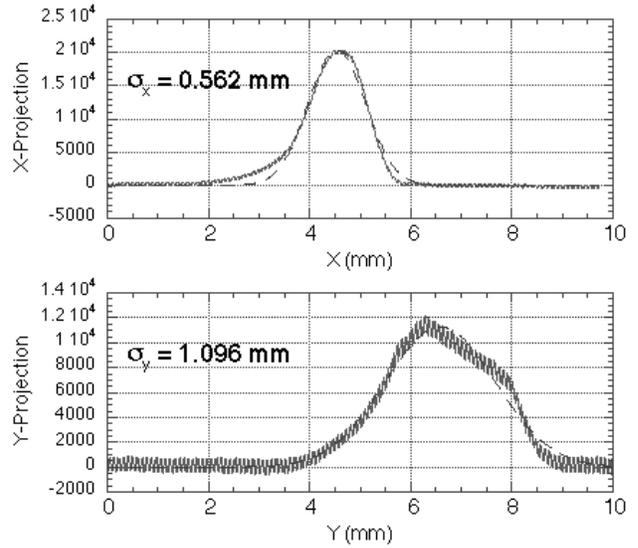

Figure 6. Projections and gaussian fits (dashed).